\documentclass[RNAAS]{aastex63}

\newcommand{\niceurl}[1]{\href{#1}{#1}}

\begin{document}

\title{Preliminary Target Selection for the DESI Quasar (QSO) Sample}

%% Note that the corresponding author command and emails has to come
%% before everything else. Also place all the emails in the \email
%% command instead of using multiple \email calls.

\author{Christophe Y\`eche}
\affiliation{IRFU, CEA, Universit\'e Paris-Saclay, F-91191 Gif-sur-Yvette, France}
\author{Nathalie Palanque-Delabrouille}
\author{Charles-Antoine Claveau}
\affiliation{IRFU, CEA, Universit\'e Paris-Saclay, F-91191 Gif-sur-Yvette, France}

% second tier
\author{David D. Brooks}
\affiliation{Department of Physics \& Astronomy, University College London, Gower Street, London, WC1E 6BT, UK}
\author{Edmond Chaussidon}
\affiliation{IRFU, CEA, Universit\'e Paris-Saclay, F-91191 Gif-sur-Yvette, France}
\author[0000-0002-4213-8783]{Tamara M. Davis}
\affiliation{School of Mathematics and Physics, University of Queensland, Brisbane, QLD 4072, Australia}
\author{Kyle S. Dawson}
\affiliation{Department of Physics and Astronomy, University of Utah, Salt Lake City, UT 84112, USA}
\author{Arjun Dey}
\affiliation{NSF’s National Optical-Infrared Astronomy Research Laboratory, 950 N. Cherry Ave., Tucson, AZ 85719}
\author[0000-0002-2611-0895]{Yutong Duan}
\affiliation{Physics Department, Boston University, Boston, MA 02215, MA}

\author[0000-0002-8281-8388]{Sarah Eftekharzadeh}
\affiliation{Department of Physics and Astronomy, The University of Utah, 115 South 1400 East, Salt Lake City, UT 84112, USA}

\author{Daniel J. Eisenstein}
\affiliation{Harvard-Smithsonian Center for Astrophysics, 60 Garden St., Cambridge, MA 02138}

\author[0000-0001-9632-0815]{Enrique Gazta\~naga}
\affiliation{Institute of Space Sciences (ICE, CSIC), 08193 Barcelona, Spain}
\affiliation{
Institut d\'~Estudis Espacials de Catalunya (IEEC), 08034 Barcelona, Spain
}

\author{Robert Kehoe}
\affiliation{Department of Physics, Southern Methodist University, 3215 Daniel Avenue, Dallas, TX 75275, USA}

\author[0000-0003-1838-8528]{Martin Landriau}
\affiliation{Lawrence Berkeley National Laboratory, 1 Cyclotron Road, Berkeley, CA 94720, USA}
\author{Dustin Lang}
\affiliation{Perimeter Institute for Theoretical Physics, Waterloo, ON N2L 2Y5, Canada}
\affiliation{Department of Physics \& Astronomy, University of Waterloo, Waterloo, ON N2L 3G1, Canada}
\author[0000-0003-1887-1018]{Michael E. Levi}
\affiliation{Lawrence Berkeley National Laboratory, 1 Cyclotron Road, Berkeley, CA 94720, USA}
\author[0000-0002-1125-7384]{Aaron M. Meisner}
\affiliation{NSF’s National Optical-Infrared Astronomy Research Laboratory, 950 N. Cherry Ave., Tucson, AZ 85719}
\author{Adam D. Myers}
\affiliation{Department of Physics \& Astronomy, University of Wyoming, 1000 E.\ University, Dept 3905, Laramie, WY 82071, USA}
\author{Jeffrey A. Newman}
\affiliation{University of Pittsburgh, 100 Allen Hall, 3941 O'Hara St., Pittsburgh, PA 15260, USA}
\author{Claire Poppett}
\affiliation{Space Sciences Laboratory at University of California, 7 Gauss Way, Berkeley, CA 94720}

\author{Francisco Prada}
\affiliation{Instituto de Astrofisica de Andaluc\'{i}a, Glorieta de la Astronom\'{i}a, s/n, E-18008 Granada, Spain}

\author[0000-0001-5999-7923]{Anand Raichoor}
\affiliation{Institute of Physics, Laboratory of Astrophysics, Ecole Polytechnique F\'{e}d\'{e}rale de Lausanne (EPFL), Observatoire de Sauverny, 1290 Versoix, Switzerland}
\author[0000-0002-5042-5088]{David J. Schlegel}
\affiliation{Lawrence Berkeley National Laboratory, 1 Cyclotron Road, Berkeley, CA 94720, USA}

\author{Michael Schubnell}
\affiliation{Department of Physics, University of Michigan, Ann Arbor, MI 48109, USA}
\author{Ryan Staten}
\affiliation{Department of Physics, Southern Methodist University, 3215 Daniel Avenue, Dallas, TX 75275, USA}

\author{Gregory Tarl\'e}
\affiliation{Department of Physics, University of Michigan, Ann Arbor, MI 48109, USA}
\author[0000-0001-5381-4372]{Rongpu Zhou}
\affiliation{Lawrence Berkeley National Laboratory, 1 Cyclotron Road, Berkeley, CA 94720, USA}
\affiliation{University of Pittsburgh, 100 Allen Hall, 3941 O'Hara St., Pittsburgh, PA 15260, USA}

%% Note that RNAAS manuscripts DO NOT have abstracts.

\begin{abstract}
The DESI survey  will measure large-scale structure using quasars as direct tracers of dark matter in the redshift range $0.9<z<2.1$ and using quasar Ly-$\alpha$ forests at $z>2.1$. We present two methods to select candidate quasars for DESI based on imaging in three optical ($g, r, z$) and two infrared ($W1, W2$) bands. The first method uses traditional color cuts and the second utilizes a machine-learning algorithm. 

%With a 260 deg$^{-2}$ fiber budget, the quasar selection will allow DESI to select at least 170 deg$^{-2}$ quasars (including at least 50 deg$^{-2}$ quasars with $z>2.1$), exceeding the project requirements.
\end{abstract}


%% Start the main body of the article. If no sections in the 
%% research note leave the \section call blank to make the title.
\section*{Introduction}

DESI~\citep{DESI2016} will precisely measure the baryon acoustic
feature imprinted on large-scale structure, as
well as the effect of redshift-space distortions on galaxy clustering.  The most distant large-scale structures will be measured using quasars (or QSOs). In the redshift range $0.9<z<2.1$, QSOs will be used as direct tracers of dark matter. At higher redshifts, DESI will analyze the foreground neutral-hydrogen absorption systems at $\lambda=1216$~\AA\ (rest frame) that comprise the Ly-$\alpha$ forest. 

DESI will primarily target QSOs using optical imaging in the $grz$ bands combined with \textit{WISE} infrared photometry in $W1$ and $W2$. Results presented here use such imaging from the DR8 release\footnote{http://legacysurvey.org/dr8/} of the DESI Legacy Imaging Surveys~\citep{Dey2019}. DESI will allocate 260 fibers deg$^{-2}$ (i.e. fibers$/\rm deg^2$)  to QSO targets, requiring at least 120 (50) of those to be QSOs at redshift $z<2.1$ ($z>2.1$).

\section*{Quasar Target Selection} 

QSOs are powered by accretion onto supermassive black holes at the centers of  galaxies. The resulting emission can greatly outshine that of the host galaxy, and the ``nuclear'' emitting regions of even the nearest QSOs are too small to resolve. QSO selection is challenging at optical wavelengths, because the colors and point-like morphologies of QSOs mimic faint blue stars. Without $u$-band imaging, selections based on the “UV excess of QSOs compared to stars cannot be implemented \citep[as for, e.g., BOSS and eBOSS;][]{Ross2012,Myers2015}. Crucially, though, QSOs are $\sim 2$ magnitudes brighter in the near-infrared at all redshifts compared to stars of similar optical magnitude and color, thus providing a powerful method to discriminate QSOs from contaminating stars.

For DESI, we have investigated two methods to select QSOs, one based on color cuts and the other on a machine-learning algorithm.  We will finalize a method during DESI Survey Validation, a four-month period prior to the start of the main survey.  We restrict both selections to objects with stellar morphology, to avoid an almost 10-fold contamination by galaxies, and we impose a depth limit of $r=22.7\,(AB)$. We also require that the targets are not in corrupted imaging pixels, nor pixels that are in the vicinity of bright stars, globular clusters, or nearby galaxies. Such ``masked" sources have \texttt{MASKBITS} of 1, 5, 6, 7, 10, 12 or 13 set in Legacy Surveys catalogs.

\begin{figure}[h!]
\begin{center}
\includegraphics[scale=0.63]{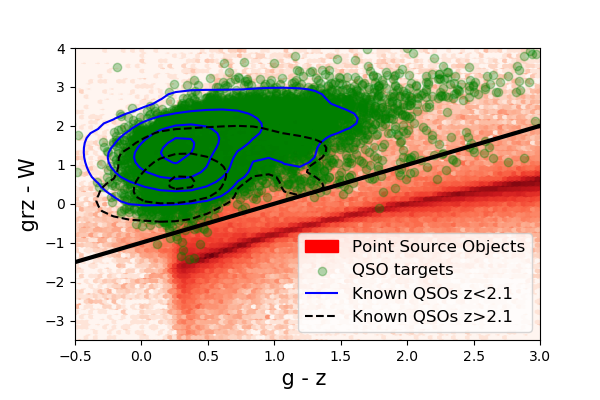}\includegraphics[scale=0.63]{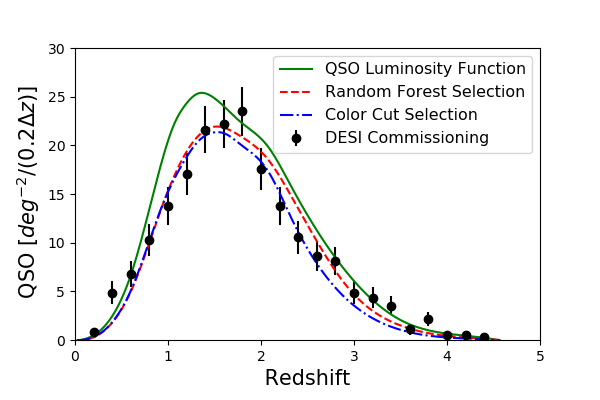}
\caption{{\bf Left:} Colors in the optical ($grz$ is a linear combination of $g$,$r$ and $z$) or  near-infrared (W is a linear combination of W1 and W2)  of  objects photometrically classified as stars  (red) or spectroscopically classified as QSOs (blue and black contours). The green circles show targets selected by the RF. The black line illustrates the ``infrared excess" color cut (sources below this line are rejected in the color-cut selection). {\bf Right:} Expected distribution of QSO redshifts from the color-cut (blue) or RF (red) selections, compared to the QSO luminosity function to $r<22.7\,(AB)$ (green). The filled circles show observations obtained during DESI commissioning.}
\label{fig:colorsQSO}
\end{center}
\end{figure}

Our color-cut selection uses $W1-W2$ to select redder sources, and $r-$W vs. $g-z$ (where $W$ is a weighted average of $W1$ and $W2$ fluxes with flux($W$)=0.75$\times$flux($W1$)+0.25$\times$flux($W2$)) to reject stars based on the ``infrared excess" of QSOs (see Figure~\ref{fig:colorsQSO}). We impose $g-r<1.3$ and $-0.4<r-z<1.1$ to avoid regions in color space almost solely populated by stars. We also restrict the magnitude range to $r>17.5$ and $grz>17$.0, where $grz$  is a weighted average of the three band fluxes (flux($grz$) = (flux($g$) + 0.8$\times$flux($r$) + 0.5$\times$flux($z)$) / 2.3)---at brighter magnitudes, QSOs are rare and stars abundant. To further reduce stellar contamination, we apply a stricter $W1-W2$ cut to sources near the stellar locus in $g-r$ vs. $r-z$. We also impose a minimum signal-to-noise ratio in both \textit{WISE} bands.

Neural-network-based algorithms implemented in BOSS \citep{Yeche2010} were found to increase QSO selection efficiency by 
$\approx 20$\% compared to color cuts. Similarly, to improve the success rate for DESI, %compared to the one obtained with the color-cut approach, 
we use a machine-learning algorithm based on Random Forests (RF). We train the RF on 230,000 known QSOs in the DESI footprint, and 210,000 ``stars'' defined as unresolved sources that are not
%in Stripe 82 (an equatorial stripe in South Galactic Cap defined by SDSS),
known QSOs and do not exhibit QSO-like variations in their SDSS light curve.  We normalized the $r$-band number counts of the stars to match the QSOs and trained the RF selection using $grzW1W2$ colors and $r$-band magnitude, varying the probability threshold with $r$.  In addition to this RF covering the whole redshift range,  we trained a specific RF to target much rarer high-redshift QSOs ($z>3.5$). To implement the RF approach, we therefore split our 260 deg$^{-2}$ fiber budget; dedicating 245 deg$^{-2}$ fibers to the all-redshift RF, and 15 deg$^{-2}$ to the high-redshift RF. This approach improves targeting for high-redshift QSOs by $\sim50$\%, with only a small loss of low-redshift QSOs.

The RF selection performs better than color cuts, particularly at $z > 2.1$ or at faint magnitudes. The expected QSO redshift distribution is shown in Figure~\ref{fig:colorsQSO}. Accounting for the completeness of both algorithms as a function of redshift and magnitude, the luminosity function of~\cite{Palanque2016} implies 195 (180) QSOs deg$^{-2}$, of which 60 (48) deg$^{-2}$ are at $z > 2.1$ for our RF (color-cut) selection. We were able to confirm these results in a $ 3.7$ deg$^2$ field observed during DESI commissioning. The spectra were visually inspected to confirm the source classification and redshift. In this field, the RF (color-cut) approaches selected $197\pm7$ ($176\pm7$) deg$^{-2}$ QSOs, including $58\pm4$ ($47\pm$4) at $z>2.1$. About $98\%$ of the QSOs selected by the color-cut approach are also contained in the RF selection.

\section*{Conclusion} 

We propose two QSO selections that meet DESI targeting requirements, and make the target catalog from the preliminary RF selection public\footnote{Available at \niceurl{https://data.desi.lbl.gov/public/ets/target/catalogs/} and detailed at \niceurl{https://desidatamodel.readthedocs.io}}. We will test these two approaches during DESI Survey Validation (SV), and choose an algorithm for the subsequent DESI main survey. During SV, we will also test complementary quasar selections such as a dropout technique to select very high-redshift quasars ($z>4.5$) and a highly homogeneous selection based on quasar variability in \textit{WISE}.
%measured with \textit{WISE} light curves. 

\acknowledgments

This research is supported by the Director, Office of Science, Office of High Energy Physics of the U.S. Department of Energy under Contract No.DE-AC02-05CH1123, and by the National Energy Research Scientific Computing Center, a DOE Office of Science User Facility under the same contract; additional support for DESI is provided by the U.S. National Science Foundation, Division of Astronomical Sciences under Contract No.AST-0950945 to the NSF's National Optical-Infrared Astronomy Research Laboratory; the Science and Technologies Facilities Council of the United Kingdom; the Gordon and Betty Moore Foundation; the Heising-Simons Foundation; the French Alternative Energies and Atomic Energy Commission (CEA); the National Council of Science and Technology of Mexico; the Ministry of Economy of Spain, and by the DESI Member Institutions.  The authors are honored to be permitted to conduct astronomical research on Iolkam Du'ag (Kitt Peak), a mountain with particular significance to the Tohono O'odham Nation.  

\bibliography{biblio}

\end{document}